# Application of a modified commercial laser mass spectrometer as a science analog of the Mars Organic Molecule Analyzer (MOMA)

Short title: Application of a MOMA analog instrument

Zachary K. Garvin[1†*], Anaïs Roussel[1,2†*], Luoth Chou[3], Marco E. Castillo[3,4], Xiang Li[3], William B. Brinckerhoff[3], and Sarah Stewart Johnson[1]

[1]*Georgetown University, Washington, D.C., USA*
[2]*INAF- Astrophysical Observatory of Arcetri, L.go E. Fermi 5, 50125 Firenze, Italy*
[3]*NASA Goddard Space Flight Center, Greenbelt, MD, USA*
[4]*Aerodyne Industries, Cape Canaveral, FL, USA*

[†]These authors contributed equally to this work and share first authorship

*Correspondence:
Zachary K. Garvin and Anaïs Roussel
3700 O Street NW
Regents Hall, Room 514
Washington, D.C. 20057
zkg3@georgetown.edu
ar1505@georgetown.edu

Published in *Frontiers in Astronomy and Space Sciences*
doi: 10.3389/fspas.2026.1781996

**Abstract**
The ESA/NASA *Rosalind Franklin* rover, planned for launch in 2028, will carry the first laser desorption ionization mass spectrometer (LDI-MS) to Mars as part of the Mars Organic Molecule Analyzer (MOMA) instrument. MOMA will contribute to the astrobiology goals of the mission through the analysis of potential organic biosignatures. Due to minimal availability of comparable equipment, laboratory analyses using similar techniques and instrumentation have been limited. Until now, the Thermo LTQ-XL platform has been used as the main analog instrument by the MOMA team despite significant differences between the instruments. In this study, we present a series of modifications that bring this commercial benchtop LDI-MS closer to MOMA operating parameters, enabling rapid testing of samples for MOMA validation experiments. We demonstrate that our instrument can detect organic standards in mineral matrices, with MS/MS enabling structural identification even in complex mixtures. Performance was additionally validated against an existing LDI-MS prototype through the comparison of spectra derived from natural samples from a Mars analog site in the Atacama Desert. Lastly, analysis of Mars analog synthetic mineral mixes highlights the capacity of the instrument to characterize both the mineralogical and organic signals in mission-relevant samples. This modified benchtop instrument will serve as a platform for collaborative research to prepare for MOMA operations, test LDI parameters, and generate pre-flight reference data in support of the mission science and astrobiology specific goals.

## Introduction

Laser desorption ionization mass spectrometry (LDI-MS) is a key analytical technique in the next generation of *in situ* astrobiology missions. This approach uses a laser to desorb and ionize analytes directly from samples and is coupled to a mass spectrometer (MS) for detailed structural identification. Critically, LDI as applied directly to natural samples offers "softer" ionization compared to conventional analytical methods (such as “harder” electron ionization, EI), enabling the characterization of non-volatile and heavier analytes (up to thousands of Da) that would otherwise be either undetectable or structurally indiscernible due to over-fragmentation.

The Mars Organic Molecule Analyzer (MOMA) instrument aboard ESA’s *Rosalind Franklin* rover will deploy LDI-MS for the first time on the Martian surface. Scheduled for launch in 2028, the rover will explore the vicinity of the landing site in the Oxia Planum province with the primary objective of seeking potential traces of ancient Martian life (Vago et al., 2017). The MOMA instrument integrates two complementary analytical approaches: (1) pyrolysis Gas Chromatography (Py-GC), building upon decades of planetary exploration heritage coupled to electron ionization (Mahaffy et al., 2012), and (2) LDI utilizing a 266 nm laser. Both methods generate ions that are swept into a linear ion trap MS for mass analysis and detection, optionally employing tandem mass spectrometry (MS/MS) for structural identification (Li et al., 2017).

Pyrolysis coupled to MS, with and without a GC, has played a central role in Martian surface exploration from the *Viking* landers in the 1970s to the Phoenix lander Thermal Evolved Gas Analyzer (TEGA) and the Sample Analysis at Mars (SAM) instrument that is currently operating on NASA's *Curiosity* rover (Klein et al., 1976; Mahaffy et al., 2012). This established technique heats scooped Martian regolith samples or drill fines to volatilize organics, which are then separated via gas chromatography before mass spectrometric identification. Wet chemistry capabilities using derivatizing agents are integrated into both SAM and MOMA to enhance detection of non-volatile molecules using py-GC-MS mode (Mahaffy et al., 2012; Goesmann et al., 2017). While SAM successfully detected trace sulfurized (Eigenbrode et al., 2018) and chlorinated (Freissinet et al., 2015) aliphatic and aromatic hydrocarbons, as well as long-chain alkanes (Freissinet et al., 2025), the search for potential biosignatures more specifically diagnostic of ancient Martian life continues. Py-GC-MS faces some analytical challenges for higher-molecular weight (m.w.) and non-volatile biosignature detection. First, both the heating and chemical derivatization processes can lead to secondary reactions that may modify potential biosignatures (Poinot and Geffroy-Rodier, 2015). Second, the widespread oxychlorine compounds, including perchlorate salts, found across the Martian surface can degrade organics by releasing $O_2$ when heated above their decomposition temperatures during pyrolysis (Navarro-González et al., 2010; Glavin et al., 2013; Ming et al., 2014). This impact can be mitigated to some extent via multistep thermal protocols (Millan et al., 2020; He et al., 2021). LDI avoids bulk sample preparation steps, rather exposing the sample surface to extremely short pulses of energy (~ns scale), resulting in desorption and "softer" ionization of (mostly nonvolatile) analytes. Additionally, Li et al. (2015) confirmed the technique's capacity to detect organic standards in mineral matrices containing as much as 1 wt.% calcium perchlorate using both commercial instrumentation and a MOMA breadboard system. Thus, LDI-MS can be utilized as a complementary technique to pyrolysis.

The related and refined technique - matrix-assisted LDI (MALDI) - employs a chemical matrix to enhance desorption and ionization while limiting analyte fragmentation. To avoid the complexities of sample preparation required by MALDI on an extremely limited rover mission, MOMA operates directly on the solid crushed sample without any preparation or contact. Several studies have demonstrated that diverse organics can be detected without matrix addition (Apicella et al., 2007; Rizzi et al., 2007). MOMA further applies the LDI protocol on samples held at Mars ambient atmospheric conditions, 4-7 Torr of predominantly $CO_2$, which is completely novel compared to classic LDI and various MALDI methods at either much lower pressure or at Earth ambient pressure (atmospheric pressure MALDI; AP-MALDI). After collecting mass spectra of sample analytes, MOMA can optionally perform tandem mass spectrometry (MS/MS) on selected parent ions, enabling identification of molecular structures through their characteristic fragmentation patterns.

In addition to MOMA, astrobiologically relevant LDI-MS instruments have been conceptualized and developed for a variety of planetary missions. The *Dragonfly* Mass Spectrometer (DraMS), which includes an LDI-MS with improvements built on top of the MOMA design, will fly to Titan as part of NASA's *Dragonfly* probe to search for life on Saturn's largest moon (Grubisic et al., 2021). An LDI-MS with an Orbitrap analyzer for improved sensitivity and mass resolution is being developed for application to missions like the proposed Enceladus Orbilander or in support of the Artemis Program (Arevalo et al., 2023). The ORganics Information Gathering INstrument (ORIGIN) is another prototype LDI-MS with specific focus on the detection of nucleobases and amino acids for application to ocean worlds like Europa (Ligterink et al., 2022; Boeren et al., 2025). Future development is also likely to adopt the use of short pulse width lasers (ps to fs) to reduce thermal-induced fragmentation, enhance ionization efficiency, and improve limits of detection (Moreno-García et al., 2016). While these instruments are valuable representations of the capabilities of LDI-MS, they have distinct differences that prevent them from serving as ideal analogs for MOMA preparation. In the context of Martian exploration, and more specifically the *Rosalind Franklin* rover mission, previous research using LDI-MS has mainly relied on commercial and prototype systems (Bishop et al., 2013; Li et al., 2015; Castillo et al., 2023). However, commercially available instruments fundamentally differ from the MOMA instrument in key parameters such as laser wavelength, energy, and fluence (energy per surface area), all of which significantly influence the desorption and ionization efficiencies, and thereby sensitivity and selectivity, of molecular species that may be intimately associated with the surfaces and/or interiors of mineral grains on a microscopic scale. Few studies have analyzed organics and mineral matrices on prototype instruments with specifications closer to those of MOMA (Li et al., 2015; Castillo et al., 2023). Some work has been performed as part of mission preparation, first using a MOMA breadboard (Li et al., 2015) and later the Engineering Test Unit (ETU) (Li et al., 2017). Nevertheless, access to these instruments remains restricted to maintain cleanliness, lifetime, and traceability to the flight instrument, which itself cannot be tested with organic-bearing analog samples.

To expand current analytical capabilities and assist in the *Rosalind Franklin* rover mission preparation, we have developed a MOMA benchtop analog instrument. In this study, we modified a commercial Thermo MALDI LTQ XL to match MOMA specifications of laser wavelength, fluence, and energy level, and tested its performance with: (1) organic standards (β-carotene, coronene, rhodamine 6G, reserpine) in isolation and spiked onto natural mineral matrices, (2) natural analog samples from the Atacama Desert, and (3) mineral matrix analogs of Martian sites (Cumberland and Oxia Planum). Our objective herein was not to fully characterize the analog samples but to validate our modified instrument and demonstrate the range of scientific analyses that can be achieved through its application. The example spectra from these samples provide a preview of future investigations and experiments with the instrument and the applicability of its results to MOMA mission-relevant analog analyses. This modified commercial platform enables high-throughput sample processing and analytical efficiency, which could guide mission operations and can facilitate interpretation of future MOMA results.

## Materials and Methods

### *Instrument description and modifications*

All described modifications were made to a Thermo Scientific MALDI LTQ XL Linear Ion Trap Mass Spectrometer operated in LDI mode (no matrix added). The initial concept for incorporating an external laser on a similar platform was based on modifications described by Korte et al. (2015) on a MALDI-LTQ-Orbitrap system. The LTQ XL was originally equipped with a MNL 100 (106-PD) nitrogen laser with a wavelength of 337 nm. To achieve MOMA parity, the laser was replaced with a Quantel Viron Nd:YAG laser (Quantel Laser, Lumibird, France) configured to emit the frequency-quadrupled wavelength of 266 nm with minor residuals of doubled (532 nm) and fundamental (1064 nm) wavelengths. Under factory settings, the maximum laser pulse energy output was 5.4 mJ (Q-switch delay = 179 µs) with a beam diameter of 2.67 mm and repetition rate of up to 20 Hz.

The LTQ laser optical path was modified for compatibility with the 266 nm wavelength and increased fluence of the Viron laser (**Figure 1**). The first dichroic mirror of the LTQ laser optical bench was replaced with a beam sampler (BSF10-UV) from Thorlabs (Newton, NJ, USA) to dump ~95% of the pulse energy of the laser, transmitting pulses in the µJ range as measured by a pyroelectric sensor (PE25-C) from Ophir Optronics (North Logan, UT, USA). The second mirror was replaced with a new dichroic mirror (#38-835) from Edmund Optics (Barrington, NJ, USA) notched to reflect 266 nm with a polished back surface for camera viewing. The original neutral density filter was removed. An additional plano-convex lens (LA4716-UV) was added to the optical path in front of the primary plano-convex focusing lens to decrease the effective focal length, thereby widening the final spot diameter on the sample plate (180 x 255 um). All other supporting optical and optomechanical components, including cage rods and mounts, were purchased from Thorlabs.

Control of the Viron laser was handled via a laser remote box interface, which enabled manual configuration via command line. Laser shots were triggered externally via the LTQ TunePlus software through TTL signals sent from the LTQ to the Viron laser via BNC connection to the diode trigger of the remote box. Though disabled, the MNL 100 laser remained electrically connected to the LTQ board via RS232 in order to maintain communications and avoid software error.

### *Sample description*

Four types of samples were selected to test the MOMA-analog instrument (**Table 1**): (1) four liquid synthetic standards, (2) a natural powder sample cleaned to remove original organics and spiked with synthetic standards, (3) a natural sample from the Atacama Desert, and (4) powdered synthetic mixes analogous to Martian sites.

Four synthetic standards—reserpine, β-carotene, coronene, and rhodamine 6G—were purchased from Sigma-Aldrich (Burlington, MA). These compounds were selected for validation and direct relevance to MOMA because of their previous use in MOMA method development, including published spectra for comparison (Li et al., 2015, 2017).

A carbonate-rich natural sample from the Green River formation (Douglas Pass, CO), was cleaned (see sample preparation details below) to remove the original organic matter and spiked with a mix of the four liquid synthetic standards listed above. The sample is mineralogically dominated by dolomite (Ca, Mg carbonate, 42 wt.%), with full mineralogy detailed in Roussel et al. (2024).

The natural sample was collected from the Green Parrot field site in the hyper-arid Yungay region of the Atacama Desert, Chile. This sample was previously analyzed by the MOMA-like Linear Ion Trap Mass Spectrometer instrument (LITMS) (Castillo et al., 2023) among other techniques.

Two synthetic mineralogical analogs were analyzed. The first, referred to here as "Cumberland," is analogous to the Cumberland sample analyzed by the instrument suite on the *Curiosity* rover at Gale Crater, which was recently shown to contain potential long chain alkanes by the *Curiosity* SAM instrument (Freissinet et al., 2025). The Chemistry and Mineralogy (CheMin) XRD analyses found the main minerals to be amorphous, plagioclase, and phyllosilicates (Bish et al., 2014) (described as "CBA" in Freissinet et al., 2020). The second mineral analog mix is the Simulant for Oxia Planum: Hydrated, Igneous, Amorphous (SOPHIA), prepared and described in Dugdale et al. (2023). SOPHIA was synthesized based on orbital data indicating that the *Rosalind Franklin* rover landing site is dominated by vermiculite, plagioclase, pyroxene, and Fe-silicate (Dugdale et al., 2023).

### *Sample preparation*

All metal tools (spatula, tweezers, aluminum foil) and glassware were cleaned of organics by baking overnight at 500°C or by solvent cleaning with methanol, dichloromethane, and hexanes (glass syringes). LDI sample plates were cleaned by sequential sonication in acetonitrile, methanol, and ultrapure water for five minutes each followed by drying under ultrapure (99.999%) $N_2$ flow.

The four liquid standards were prepared from stock powders to a concentration of 10 ng/µL in dichloromethane. Three spots of 1 µL were deposited on the LDI plate and air dried in a fume hood.

The carbonate-rich natural sample was baked at 500°C overnight to remove endogenous organic material (around 7 wt. % TOC, see Roussel et al., 2024) and spiked with a dichloromethane (DCM) solution containing the four synthetic standards. The stock DCM solution was prepared with 0.4 µg/µL of reserpine and β-carotene, and 0.8 ng/µL of coronene and rhodamine 6G. A volume of 50 µL of solution was added to 40 µg of sample powder, for a final concentration of 500 ng std / mg powder (500 ppm) for reserpine and β-carotene, and 1 ng std / mg powder (1 ppm) of coronene and rhodamine 6G. To improve homogenization of the standards to the mineral matrix, the spiked powder was sonicated in a fume hood for 5 min in a closed vial, then 15 min in an open vial while slowly air drying.

All powdered samples (carbonate spiked with organics, Atacama, Cumberland, and SOPHIA) were deposited onto the LDI plate using the same method. As the LDI plate sits vertically in the sample chamber during analyses, the powder must be affixed to the plate by double-sided tape (3M, adhesive transfer tape 966). A few grains of the powder are deposited on the tape and gently pressed down with a clean metal spatula to secure any loose grains that could fall and contaminate the sample chamber. The powder is then briefly placed under a flow of clean $N_2$ to remove excess or unattached grains. Analysis of blank tape without the addition of powder shows a low level of background signal that does not interfere with the interpretation of the spectra herein (see **Supplementary Figure S1)**.

### *Mass Spectrometry*

After loading the samples, the sample plate is inserted into the LTQ and evacuated to 75 mTorr of $N_2$ in the source region to limit fragmentation and reduce ion transfer time during laser-induced ionization (Strupat et al., 2009). The laser beam is directed into the source region and focused on the sample plate at an incident angle of 32° relative to the laser beam (compared to approximately 45° for MOMA, Siljeström et al., 2021). The sample plate is moved in x or y increments of 250 μm to adjust the laser position on the samples. Ions are directed into the mass analyzer region through a series of ion optics, including quadrupolar and octupolar rod assemblies serving as ion guides.

LDI-MS was performed in positive ion mode within a scan range of 50-2000 *m/z* and a resolution (full-width half maximum; FWHM) of 0.7, compared to the MOMA scan range of 50-1000 *m/z* and resolution (FWHM) of ≤2 (Goesmann et al., 2017). Up to three laser pulses were fired per scan. A maximum of 5 scans were acquired at each position before moving the plate to the next location. Spectra were recorded by averaging across scans to produce a final representative spectrum. Laser energy was adjusted between data acquisition sequences as needed by varying the Viron Q-switch delay according to an energy calibration (see Results and Discussion and **Figure 2**). Automatic Gain Control (AGC) and Automatic Spectrum Filtering (ASF) of the LTQ were switched off to maintain manual control of energy for the Viron laser and manual curation of spectra, respectively. Replicate samples were run for validation to assess sample heterogeneity and potential contamination.

Tandem mass spectrometry (MS/MS) was performed on samples to further characterize target masses of interest. MS/MS parameters were set within the Tune Plus v2.5.0 software. Targets were selected manually by parent mass of interest. The normalized collision-induced dissociation (CID) energy was set individually for targeted compounds/fragments to achieve ideal fragmentation for each parent mass. The isolation width was set to 3 *m/z*.

### *Data Analysis*
We used the Thermo Qual Browser v2.0.7 to identify parent ions (MS) and confirmed their structures using their fragmentation patterns with MS/MS. The MS and MS/MS spectra were averaged across 30 and 10 scans respectively to improve the signal intensity and cover a larger area of sample, thereby reducing the heterogeneity of the sample.

## Results and Discussion
### *Laser parameters*
The modified laser parameters and a comparison among related instruments are presented in **Table 2**. The energy range of the Viron laser was calibrated with a pyroelectric sensor across Q-switch (QS) delay values from 179 to 210 μs. Due to initial laser pulses exhibiting higher energy and greater variability between pulses, the first 10 pulses after initiating laser firing were excluded from the energy calibration (**Supplementary Figure S2**) and sample analyses. As the MALDI plate moves between laser shots, the spectra we show here are from spots on the sample that were not exposed to these initial high-energy laser pulses. Laser pulse energies were observed to attenuate slowly; to assess energy stabilization, the laser was continuously fired and energy measured at intermittent times over 120 min (**Supplementary Figure S3**). Sufficiently stable energy per pulse was reached after approximately 30 min. Thus, energy calibration was performed after a 30 min warm-up period, pulsing with the laser shutter closed at QS delay set to 200 μs. To ensure consistent desorption and ionization in subsequent analyses, all mass spectra are based on scans taken after a 30 min warm-up, and only spectra collected after the first 10 pulses of each analysis once energy has stabilized are utilized for downstream spectral averaging. This corrected calibration corresponds to energy levels from 7.4 ± 1.8 to 75.3 ± 7.3 μJ (**Figure 2, Supplementary Table S1**). Laser pulse energy below the stable limit of detection of the energy meter was estimated based on a $2^{nd}$ order polynomial fit to the calibration data ($R^2$ = 0.998), revealing that the laser reaches values as low as 0.9 μJ at a QS delay of 220 μs.

The measured energy range overlaps with the reported range of the MOMA laser (~13-130 μJ) (Goesmann et al., 2017). However, the smaller spot size of the focused Viron laser beam (~53× smaller area than MOMA) resulted in a higher total fluence compared to MOMA. To address this discrepancy as well as to increase the target surface area to desorb more analyte per laser shot, an additional plano-convex lens was added to the optical path in order to increase the final spot size. Gaussian beam propagation calculations indicated a lens with focal length of 752.6 mm placed ~350 mm before the primary plano-convex focusing lens would achieve a spot size roughly equivalent to MOMA. The final laser spot size after this adjustment was expanded by a factor of 10 from 50 x 90 μm to 180 x 255 μm, corresponding to a spot size ~5× smaller than MOMA (400 × 600 μm; **Figure 3**). The laser spot dimensions were estimated using a layer of α-Cyano-4-hydroxycinnamic acid (CHCA, a matrix classically used in MALDI) due to its enhanced ionization properties and clear visual evidence of the laser spot after firing. 1 uL of CHCA dissolved in acetonitrile was deposited directly on the plate and air-dried in the fume hood (quantification of CHCA was unnecessary here as its purpose was only visual).

With the increased laser spot size, the QS delay 198 to 216 μs covers the MOMA operational specifications with a fluence spanning from 0.007 to 0.070 J $cm^{-2}$, and an irradiance from 0.4 to 34.8 MW $cm^{-2}$ (**Figure 2, Table 2, Supplementary Table S1, Supplementary Figure 4**).

***Spiked natural minerals***

Liquid organic standards were used to validate the performance of our modified LDI-MS (**Supplementary Figure S5**) and to collect MS/MS fragmentation patterns to confirm identifications (illustrated in **Supplementary Figure S6**). We successfully detected the parent ions of the spiked coronene ($M^+$ at *m/z* 300), reserpine ($M^+$ at *m/z* 608), β-carotene ($M^+$ at *m/z* 536), and rhodamine 6G ($M^+$ at *m/z* 444) on the ashed Green River sample (**Figure 4**).

Our results confirm the detection of trace amounts (1 to 500 ppm) of known organics on complex natural mineral matrices using our MOMA-like LDI-MS instrument in MS and MS/MS modes. Though limits of detection will vary depending on target molecule, mineral matrix, and laser parameters, these measurements suggest comparable sensitivity to the MOMA breadboard and are within mission science goals, reaching <1 pmol $mm^{-2}$ levels of detection of rhodamine 6G (Brinckerhoff et al., 2013; Goesmann et al., 2017). Such tests also form an important element of the natural sample data library for LDI-MS. Most analyses of samples containing a physical mixture of organic compounds in a mineral matrix produce complex spectra with peak distributions and intensities related to the dissociation of mineral phases and the fluence-dependent efficiency of organic desorption/ionization in that context. As such, experiments involving data recorded at multiple laser pulse energies combined with MS/MS targeting of diagnostic species are generally required to make confident assignments. With unknowns, this process is strongly supported by such reference sample library spectra. Further experiments with this instrument will assist in expanding such a library, including (1) testing spiked organics detection on more diverse mineralogies (i.e., clays, sulfates), as well as (2) using organics more relevant for astrobiology (i.e., lipids, amino acids, nucleotides).

***Natural Mars analog sample***

To compare the performance of our modified LTQ against results from a similar instrument, we conducted a depth series analysis of Atacama soil samples previously examined in a set of field and laboratory tests with the Linear Ion Trap Mass Spectrometer (LITMS) instrument at NASA Goddard Space Flight Center. The LITMS design follows MOMA but features enhanced functionality such as detection of both positive and negative ions (Castillo et al., 2023). LITMS utilizes the same Viron laser with a different optical path setup, resulting in a different spot size and fluence/energy range than the LTQ (**Table 2**).

Comparison of spectra derived from the two instruments reveals similarities in major identifiable peaks (our LTQ data in **Figure 5**, LITMS data in Castillo et al., 2023). Among the most abundant masses, peaks at *m/z* 299 and 315 are common to all three depths from the LTQ analyses. Similar corresponding peaks are present in the shallow depths of the LITMS data but are not noticeably abundant at the 50 cm depth. These peaks are presumed to represent a primary component of the sample mineralogy based on the relatively low mass and the likely loss of oxygen inferred from the 16 *m/z* difference. MSMS analyses appear to support this interpretation, showing M-18 peaks suggestive of water loss (**Supplementary Figure S7**). A peak at *m/z* 113, also hypothesized to be mineralogical, is identifiable at all three depths from the LTQ data, though is most prominent in the surface sample. A potential corresponding peak at *m/z* 112 is observed in the shallow sample in LITMS (Castillo et al., 2023), while *m/z* 113 is the most abundant peak at 50 cm depth. Due to observed minor drift of the mass scale and peak shape fluctuations in the smaller LITMS ion trap, as well as due to both $M^+$ and $MH^+$ molecular ionization observed in LDI of mineral samples, it is possible that dominant peaks variously assigned to *m/z* 299 or 300, m/z 315 or 316, and *m/z* 112 or 113 do reference the same structures in these samples. However, additional mass calibration and ideally MS/MS data would be needed for confirmation, not yet available for the LITMS results. Future analyses will focus on characterizing peaks from these samples to provide a complimentary dataset to assist in the interpretation of data from flight-like prototypes.

Ions of higher masses are also observed in the LTQ results. Clusters of peaks around *m/z* 577, 761, and 945 are present in all three depths at varying abundances (**Figure 5**). The consistent separation of these peaks by 184 *m/z* could indicate a polymeric relationship or the loss/addition of a common lipid headgroup. Further investigation via MSMS reveals primary fragments at M-112 for each of these target masses, clustered with other peaks of ±1 *m/z* (**Supplementary Figure S8**). This additional fragmentation evidence could point to a loss of a hydrocarbon component like octene (*m/z* 112), lending additional support to the assignment of this series of peaks as lipids (e.g., diacylglycerols) with fatty acid tails. However, definitive identification of specific lipid molecules would require other complimentary methods.

An overall decrease in background signal is observable in the LTQ data with increasing depth. The LTQ signal across all depths is most abundant in the lower mass range, likely indicating a stronger inorganic signal. LITMS data was more evenly distributed in peak intensity from *m/z* 100 to 500, potentially due to the LITMS analyses being performed with a smaller mass range (up to *m/z* 500), thereby eliminating contribution from the fragmentation of larger ions that were not sent to the ion trap. Both the similar low mass peaks and the distributional discrepancies between the LTQ and LITMS analyses could also be reflective of a large difference in fluence. Due to the focused spot size of the LTQ, the laser is estimated to be ~100× higher fluence than LITMS (see **Table 2**). Thus, despite operating the laser at a lower energy than the LITMS study (~16 vs. ~80 μJ), the elevated energy density of the LTQ setup could have enabled the detection of the mineralogical peaks that were only detectable by LITMS at a higher energy. The overall shift to lower masses could also be due to increased fragmentation of nonvolatile organics at higher masses or a stronger inorganic signal.

***Mars analog synthetic mineral mix***

The results for the two mineralogical analogs acquired with a laser fluence of 0.070 J $cm^{-2}$ (MOMA maximum) are presented in **Figure 6**. SOPHIA's most abundant peaks are lower masses (between 130 and 300 Da), likely predominantly reflecting the major mineral composition (**Figure 6A**) (Dugdale et al., 2023). Distinct higher m/z and less intense peaks likely indicate organic compounds present in the mix, preserved or adsorbed in the individual powders. While the SOPHIA mix was prepared in organically clean conditions, minimizing surficial contamination, mineral particulates and crystallites are not subjected to acid digestion or aggressive organic removal procedures (to maintain sample integrity) and thus may host incorporated organic species at fine scales that have been observed in LDI-MS analyses (Dugdale et al., 2023). Similarly, the Cumberland analog results (**Figure 6B**) also exhibit peak patterns consistent with a complex mineral matrix and minor or trace organic components. Future work in the context of mineralogical analogs includes: (1) analyzing individual minerals from these mixtures to identify each peak within the mix's spectra and (2) spiking the sample (whole or phase separated) with known organic standards, following the procedure used for the Green River sample, to understand ionization behavior and biases in complex mineralogical matrices, directly relevant for Martian exploration.

***Notable differences from MOMA***

While the modifications to the MALDI LTQ XL system have produced a system closely replicating MOMA functionality, providing an important laboratory capability for analog studies, some key discrepancies remain. Foremost among these is that in MOMA, LDI occurs at Mars ambient conditions, 5-7 Torr of primarily $CO_2$, whereas the LTQ sample plate is maintained at less than 100 mTorr of $N_2$. Some differences in the distribution of induced molecular and fragment ions between the higher pressure and density of the MOMA system and the lower pressure and density of the "more MALDI-like" LTQ system are thus expected, depending on the sample composition. Furthermore, in the case of MOMA, ions are guided directly into the ion trap via simple capillary via a discontinuous atmospheric pressure interface (DAPI) following Gao et al. (2008), utilizing a pulsed aperture valve and associated dynamic pressure environment. Essentially, the pressure in the ion trap is rising during ion formation and transport and falling during ion trap cooling and any MS/MS scanning. Whereas the LTQ includes a series of additional ion optics (quadrupole and octupole) preceding the ion trap, which is differentially pumped to a static low pressure of He. These differences may result in different patterns of reactions, recombination/neutralization, and fragmentation during transport. The MOMA mass spectrometer is also 25% the size of the LTQ, impacting the storage capacity and manipulation of ions (Li et al., 2017).

Additional differences relate to the control of ions during and after the LDI process, which could affect result comparability. MOMA utilizes a form of automatic gain control (AGC) to optimize laser energy for maximum ion signal (Goesmann et al., 2017). For our modified LTQ, AGC is disabled, and laser energy is controlled manually due to communication limitations and to avoid losing ions during AGC laser shots. The MS/MS methodology differs somewhat between the two instruments. The collision gas of the LTQ is He rather than the ambient Mars atmosphere (predominantly $CO_2$) for MOMA. Future work could incorporate the use of $CO_2$ gas for CID collision and in the sample chamber. The CID energy parameter for MS/MS is also not directly comparable between instruments as the reported CID energy of the LTQ is normalized by the manufacturer without precise reporting of the applied voltages to the trap (Lopez et al., 1999). While these differences necessitate careful cross-calibration, interpretation, and periodic comparison of data between the two instruments, our results to-date are promising for the great utility of the modified LTQ to serve as an analog instrument to assist in the preparation for the upcoming mission and potentially Mars science data analysis. This benchtop analog particularly enables high-throughput MOMA-like experiments on samples containing organics in order to gain insight into important LDI and MS/MS processes with Mars analog samples without burdening or risking unique MOMA flight-like hardware.

## Conclusions

We have successfully modified a commercial benchtop Thermo MALDI LTQ XL linear ion trap mass spectrometer to achieve a high degree of parity with the MOMA instrument design and functionality. The original system laser was replaced to match the 266 nm wavelength of the MOMA laser to better replicate the MOMA desorption and ionization process. Additional modifications to the laser optical path and communications yielded an energy range and spot size consistent with MOMA specifications, permitting a fully adjustable fluence within MOMA operating parameters and enabling automatic laser firing synced with data acquisition via communication with the mass spectrometer. The MOMA mass spectrometer itself is a scaled-down variant of the LTQ model, making the linear ion trap design of the LTQ the closest commercial equivalent to MOMA. With these modifications, the instrument currently stands as the only operational analog instrument for the LDI component of MOMA and the closest to its functional parameters other than the ETU model, which must be used sparingly to preserve its state for the mission.

Initial instrument testing demonstrated successful detection of nonvolatile organic standards, both in solution and within a mineral matrix, as will be analyzed on Mars. Targeted MS/MS analyses further enabled identification of these organics in a mixture, replicating a key functionality of MOMA for detecting biosignatures at low concentrations and confirming their structures within complex samples. Analysis of low biomass samples from the hyper-arid Atacama Desert demonstrated the capability of the modified LTQ to process natural samples collected from Mars analog environments, yielding results comparable to those obtained using other established flight capable LDI instruments. For direct relevance to Mars, we analyzed synthetic mineral mixes analogous to the known compositions of Oxia Planum and Cumberland. While the organic content of the analog minerals cannot be used as an indicator for Martian biosignatures due to their terrestrial origins and methods of preparation, the mineral mixes provide a mineral background signature for comparison with future MOMA data during mission operation. The data from the modified LTQ provides information about the ionizable components of the minerals expected to be present in these selected Mars sites, establishing a critical library of reference spectra for MOMA.

Together, these results validate the functionality of our modifications to the LTQ and display the capacity of the instrument to perform MOMA-like analyses. As the community prepares for the launch of the Rosalind Franklin rover to Oxia Planum in 2028, preparation for the science objectives will be critical to maximize scientific return and enable efficient and effective use of the instrumentation during the mission lifetime. MOMA will serve as the next generation mass spectrometer being sent to Mars and will play a crucial role in geochemical characterization and organic biosignature detection. Further use of the modified LTQ as an analog instrument for MOMA would help to prepare and plan for actual flight experiments and data analysis. The instrument can effectively run numerous MOMA experiments rapidly in succession due to the automatic laser positioning control across and among samples. Thus, the modified LTQ provides a unique opportunity for high throughput testing of relevant samples and experimental protocols. The instrument offers a platform for future collaborative work to generate fundamental datasets that will lay the groundwork for MOMA to achieve its astrobiology goals.

## Acknowledgments

We thank Professor Young-Jin Lee for inspiring and catalyzing our journey in modifying the LTQ. We thank Dr. Amy Dugdale, Dr. Nisha Ramkissoon, and collaborators for providing the SOPHIA sample. We are immensely grateful to Dr. Ed Van Keuren at Georgetown University for his invaluable assistance with laser optics. We also thank Dr. Christine Knudson for providing the Cumberland analog sample and Dr. Paul Mahaffy for countless illuminating discussions on mass spectrometry and beyond.

**Funding Statement**
This work was supported by NASA award 80NSSC18K1140. Additional support to Z.K.G. was provided by the Georgetown University Science, Technology, and International Affairs (STIA) Postdoctoral Fellowship.

**Data Availability Statement**
All LDI-MS raw files are available in the following Figshare repository: https://figshare.com/s/fb161d42e6e2fd8a2de4

**Author Contributions**
S.S.J., W.B.B., and L.C. formulated the instrument modification concept. L.C. refurbished the commercial instrument and replaced the original laser. A.R. and Z.K.G. completed the laser path modification, energy calibration, and spot size optimization, as well as the instrument validation on samples. M.E.C., X.L., and W.B.B. advised on set-up modifications, optimization, and data interpretation. A.R. and Z.K.G. wrote the original manuscript, and all co-authors reviewed and edited the article.

**Competing Interests**
The authors declare no competing financial interests.

**List of abbreviations**
AGC: Automatic Gain Control
ASF: Automatic Spectrum Filtering
CHCA: α-Cyano-4-hydroxycinnamic acid
CID: Collision Induced Energy
DCM: Dichloromethane
GC-MS: Gas Chromatograph – Mass Spectrometer
LIT: Linear Ion Trap
LDI-MS: Laser Desorption/Ionization – Mass Spectrometer
LITMS: Linear Ion Trap Mass Spectrometer
MALDI: Matrix Assisted Laser Desorption Ionization
MOMA: Mars Organic Molecule Analyzer
SOPHIA: Simulant for Oxia Planum: Hydrated, Igneous, Amorphous
Py: Pyrolysis

## Tables

**Table 1.** Sample information: type, role in this study, and previous descriptions in the literature.

| Sample name | Sample description | Role | Previous publications |
|---|---|---|---|
| Liquid organics | Reserpine, b-carotene, coronene, rhodamine 6G | Test ionization with new instrument (MS), confirm fragmentation pattern (MSMS) | Li et al., 2015; Li et al., 2017 |
| Green River spiked | Ashed dolomite-rich natural sample, spiked with organic standards | Test known organic detection in complex natural sample matrix. | Roussel et al., 2024 |
| Atacama | Natural soil sample | Test a complex natural sample, and compare results to LD-MS flight test instrument | Castillo et al., 2023 (named "19A" surface, 20 cm, and 50 cm) |
| Cumberland | Mineralogical analog mixes | Test mineralogical signal comparable to Martian sites | Freissinet et al., 2020 |
| SOPHIA | | | Dugdale et al., 2023 |

**Table 2.** Parameter comparisons among LD-MS mission and commercial instruments. GU LDMS (described in this study), MOMA (Rosalind Franklin rover), LTQ XL (original commercial instrument), and LITMS (mission prototype instrument).

| Parameter | GU LDMS | MOMA | LTQ XL | LITMS |
|---|---|---|---|---|
| **Energy range** *(µJ)* | 1 to 75 | 13 to 130 | 0.1 to 100 | 25 to 250 |
| **Spot diameter** *(µm)* | 180 x 255 | 400 x 600 | 50 x 90 | ~300 |
| **Spot area** *($cm^2$)* | 3.6E-04 | 2.0E-03 | 1.6E-04 | 7.1E-04 |
| **Fluence range** *($J\ cm^{-2}$)* | 0.002 to 0.209 | 0.007 to 0.069 | 0.001 to 0.637 | 0.035 to 0.354 |
| **Pulse width** *(ns)* | 6 | 1 | 3 | 6 |
| **Irradiance range** *($MW\ cm^{-2}$)* | 0.4 to 34.8 | 7.0 to 69.0 | 0.3 to 212.3 | 5.8 to 59.0 |

## Figures

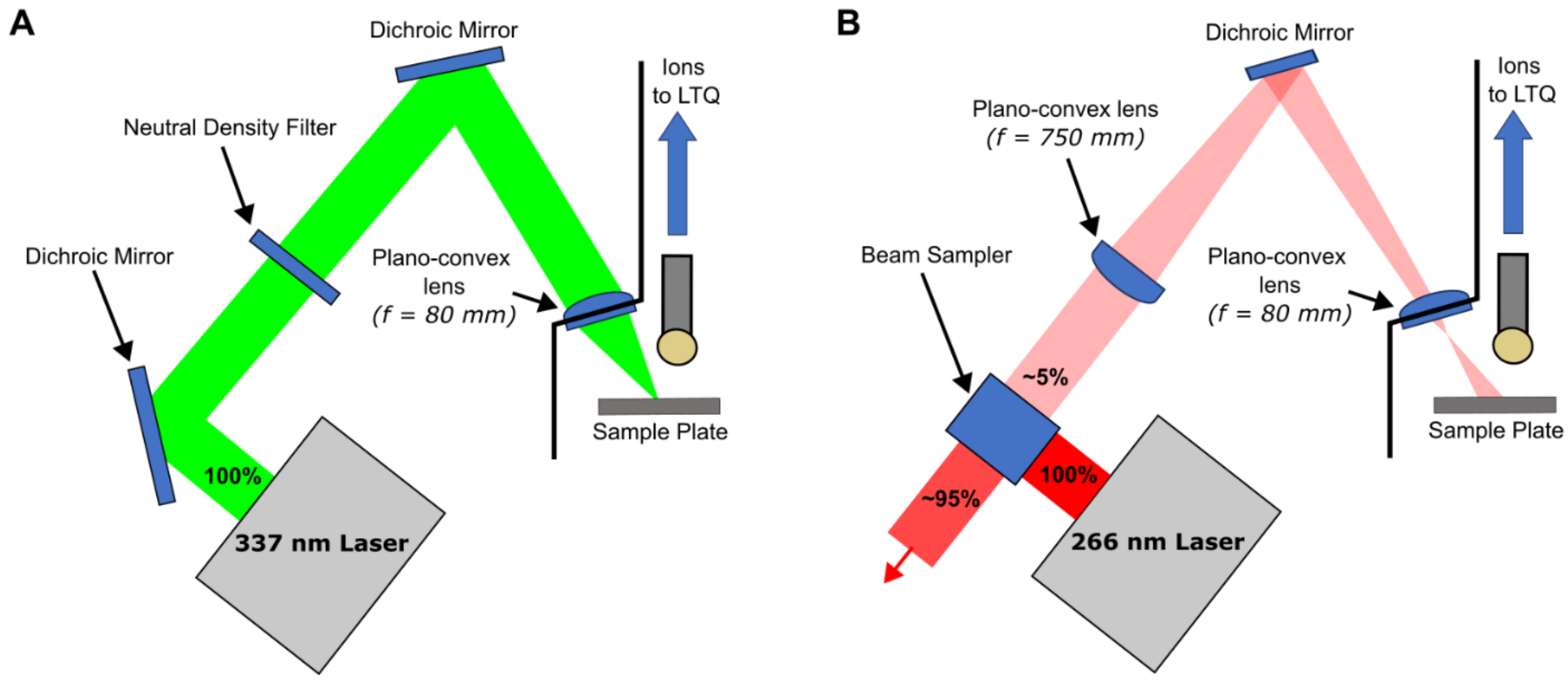

**Figure 1.** Schematic of (A) the original LTQ 337 nm laser optical path and (B) the modified 266 nm laser optical path. In the modified path, the initial laser beam is split by a beam sampler, dumping ~95% of the total energy. The 5% "sampled" beam passes through a plano-convex lens (f = 750 mm) and is reflected toward the instrument via a dichroic mirror. The beam then passes through the built-in focusing lens (f = 80 mm) of the LTQ to enter the source region where it contacts the sample plate. Ions are guided through a quadrupole and into the mass spectrometer.

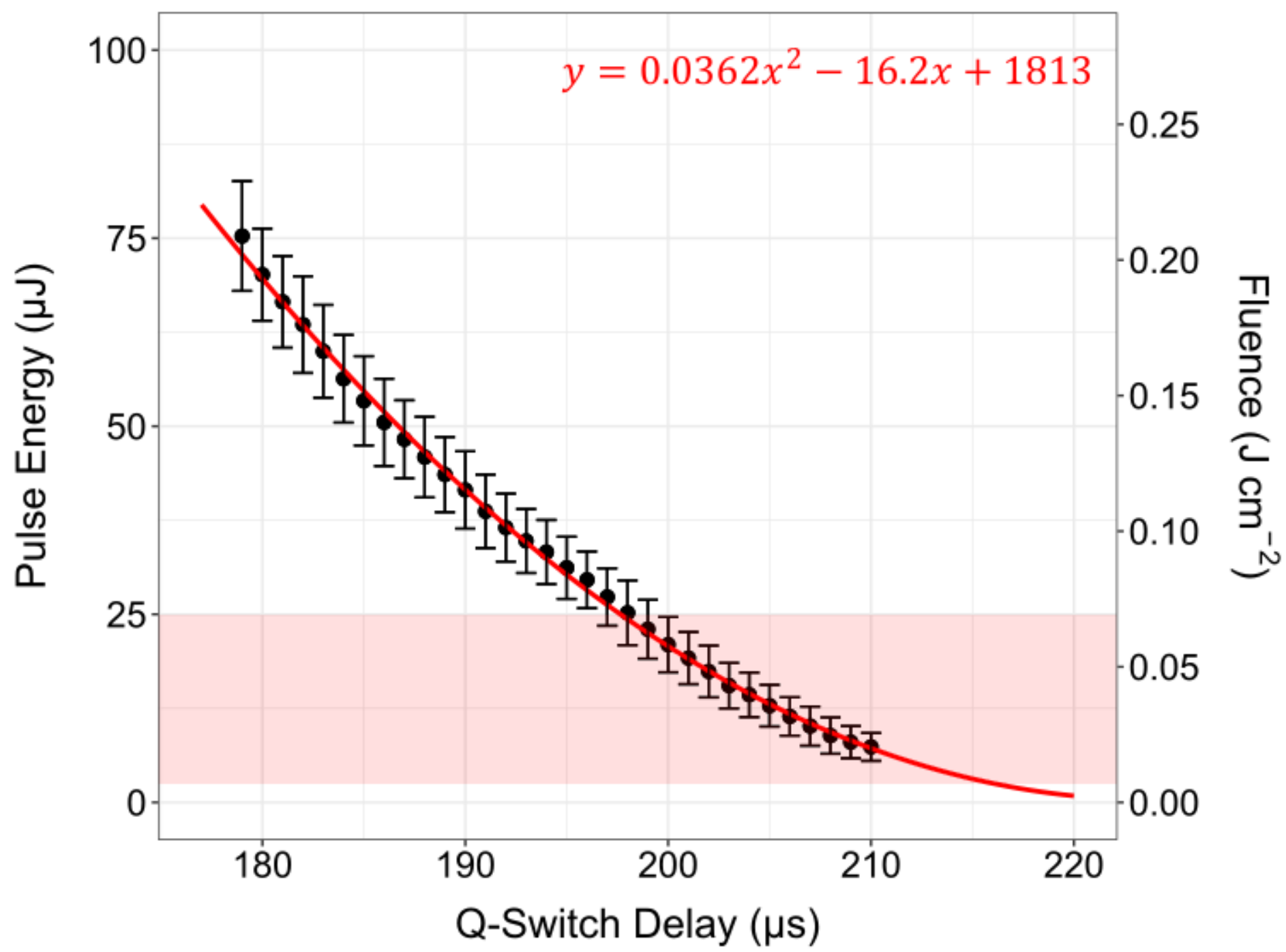

**Figure 2.** Calibration of pulse energy (μJ) across Q-Switch delay. Secondary y-axis on the right indicates the corresponding laser fluence (J cm$^{-2}$) for the final modified beam size (180 x 255 μm). The curve represents the average of two series of data collected after 30 and 60 minutes of laser warm-up, with removal of the first laser 10 pulses, and fits a second order polynomial regression (adjusted $R^2 = 0.998$). The corresponding MOMA fluence range is highlighted in red. Data is listed in Supplementary Table S1.

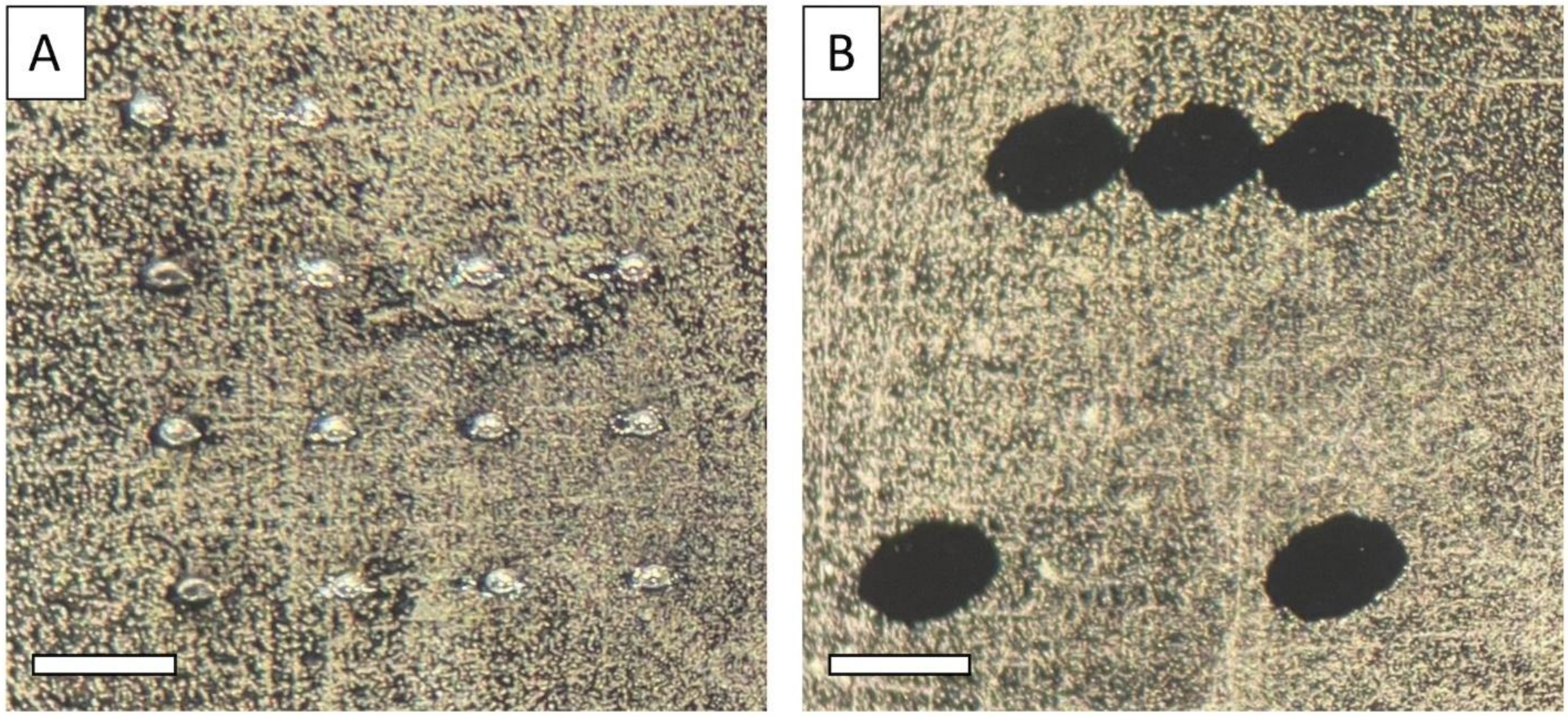

**Figure 3.** Laser beam spot dimension on a thin layer of CHCA matrix (A) before (50 x 90 µm) and (B) after (180 x 255 µm) laser path optimization. White scale bar = 250 µm. To maintain consistency with MOMA functionality, CHCA was not used in any experiments in this study. Here, it served only to characterize the laser beam spot due to its high ionization efficiency and clear visual residue on the sample plate for size measurement.

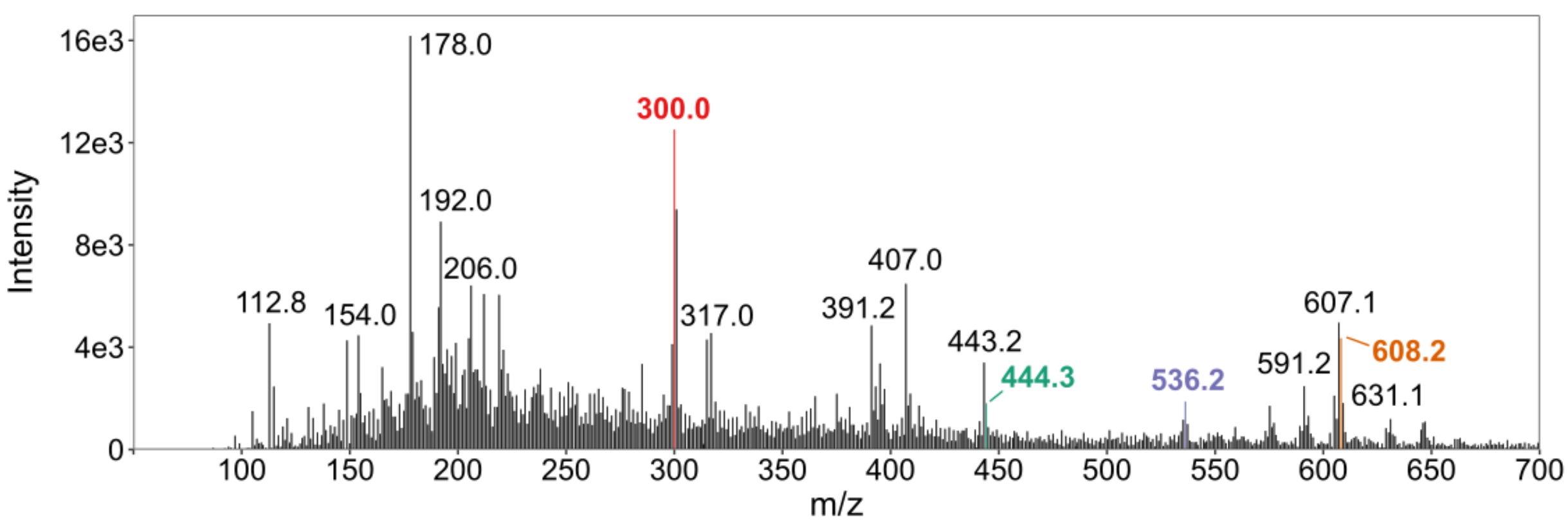

**Figure 4**. Modified LTQ XL mass spectrum of spiked organic mix on Green River ashed powder. Spiked concentrations are 500 ng/mg for reserpine ($M^+$ at m/z 608; orange) and β-carotene ($M^+$ at m/z 536; purple), and 1 ng/mg for rhodamine 6G ($M^+$ at m/z 444; green) and coronene ($M^+$ at m/z 300; red). Parent ion masses are indicated in bold. Laser fluence at 0.007 J $cm^{-2}$, comparable to the MOMA minimum fluence (See Supplementary Table S1).

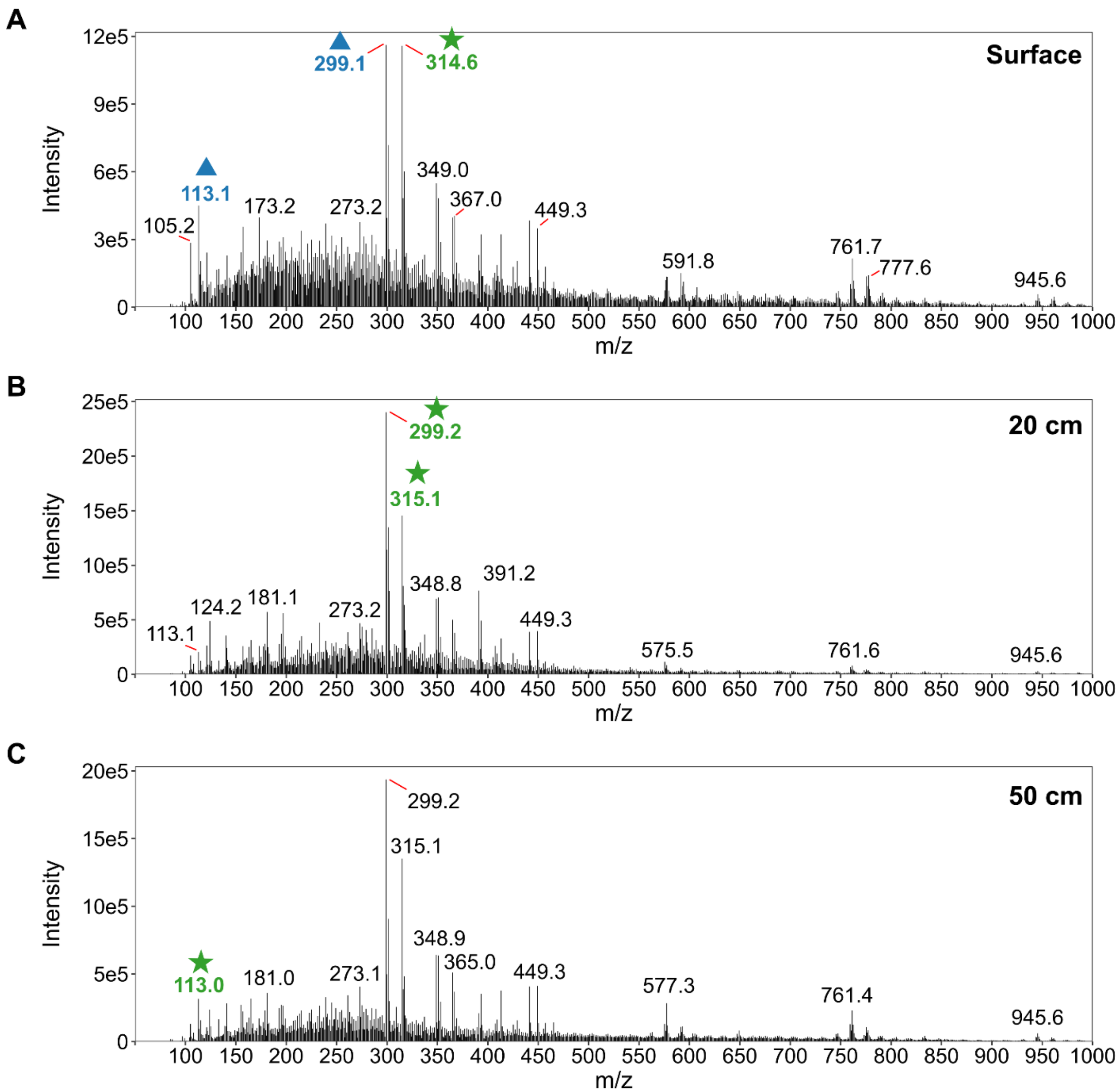

**Figure 5.** Modified LTQ XL mass spectra of Atacama soil samples from (A) surface, (B) 20 cm depth, and (C) 50 cm depth at the Green Parrot site. Samples were originally collected and analyzed on LITMS by Castillo et al. (2023). Peaks marked with shapes indicate overlap with LITMS results at corresponding depths, either exact mass (green stars) or drifted ±1 m/z (blue triangles). Laser fluence at 0.043 J $cm^{-2}$, comparable to a MOMA mid-range fluence (See Supplementary Table S1). Select MSMS spectra of targets m/z = 299, 315, 577, 761, and 945 are available in Supplementary Figures 7 and 8.

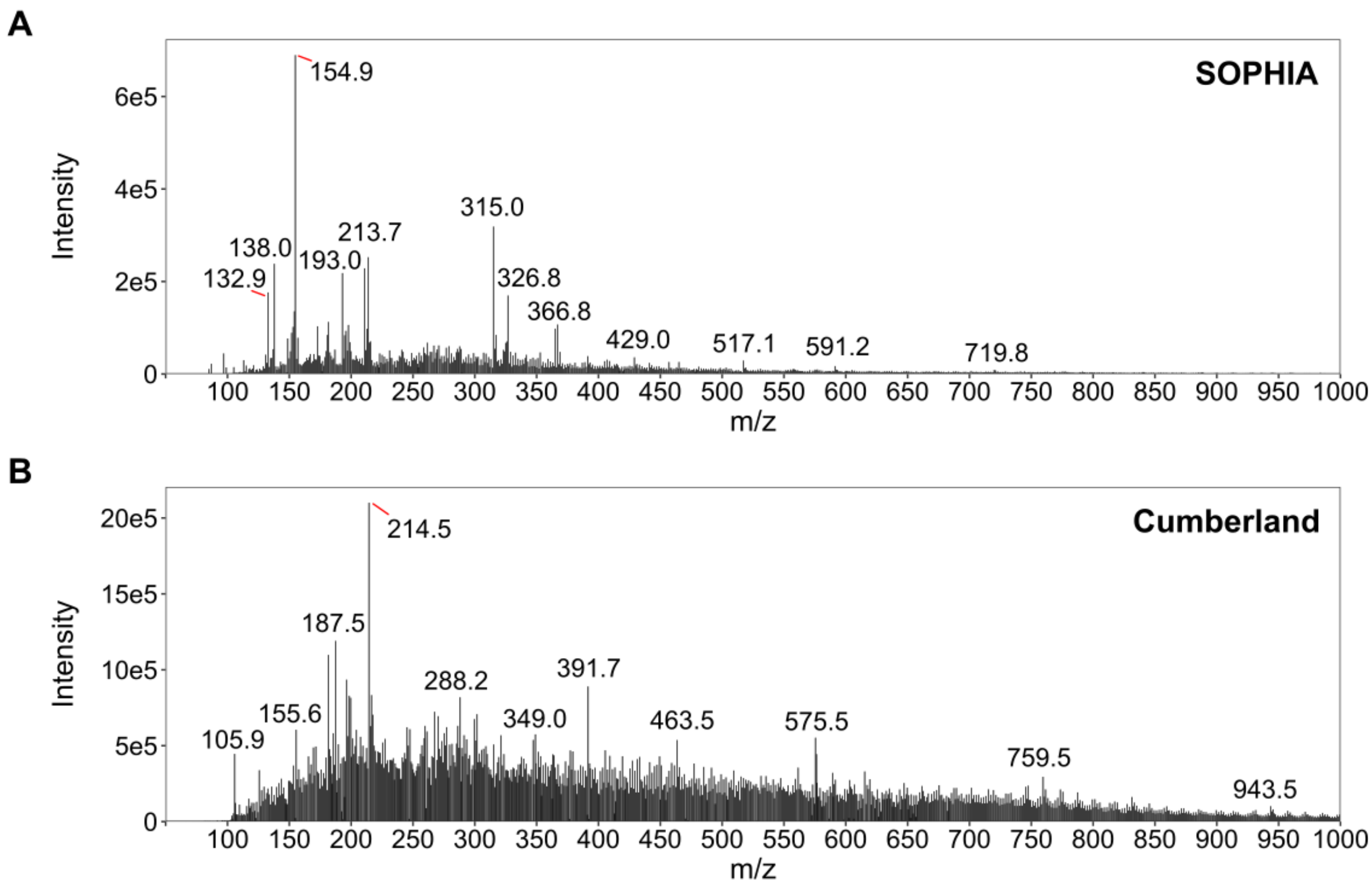

**Figure 6.** Mass spectra of two mineralogical Martian analogs: (A) SOPHIA and (B) Cumberland mix. Laser fluence at 0.070 $J.cm^{-2}$, comparable to MOMA maximum fluence (See Supplementary Table S1).

## Supplementary Material

### *Figures*

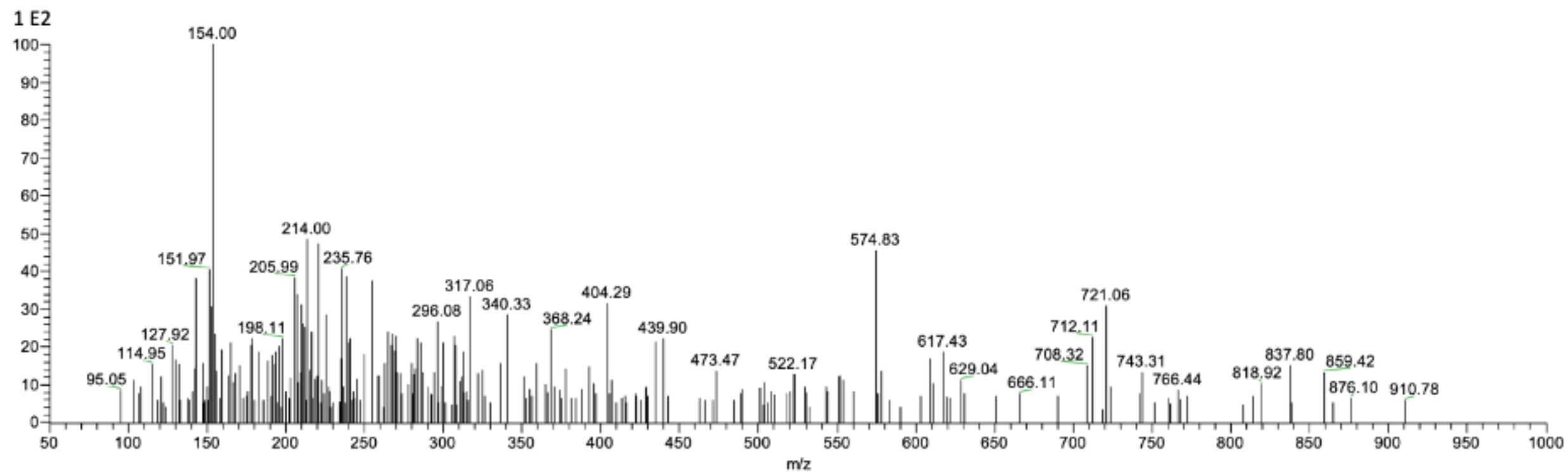

**Supplementary Figure S1.** Mass spectrum of double-sided tape background signal. Laser fluence at 0.043 J $cm^{-2}$, comparable to MOMA mid-range fluence (See Supplementary Table S1).

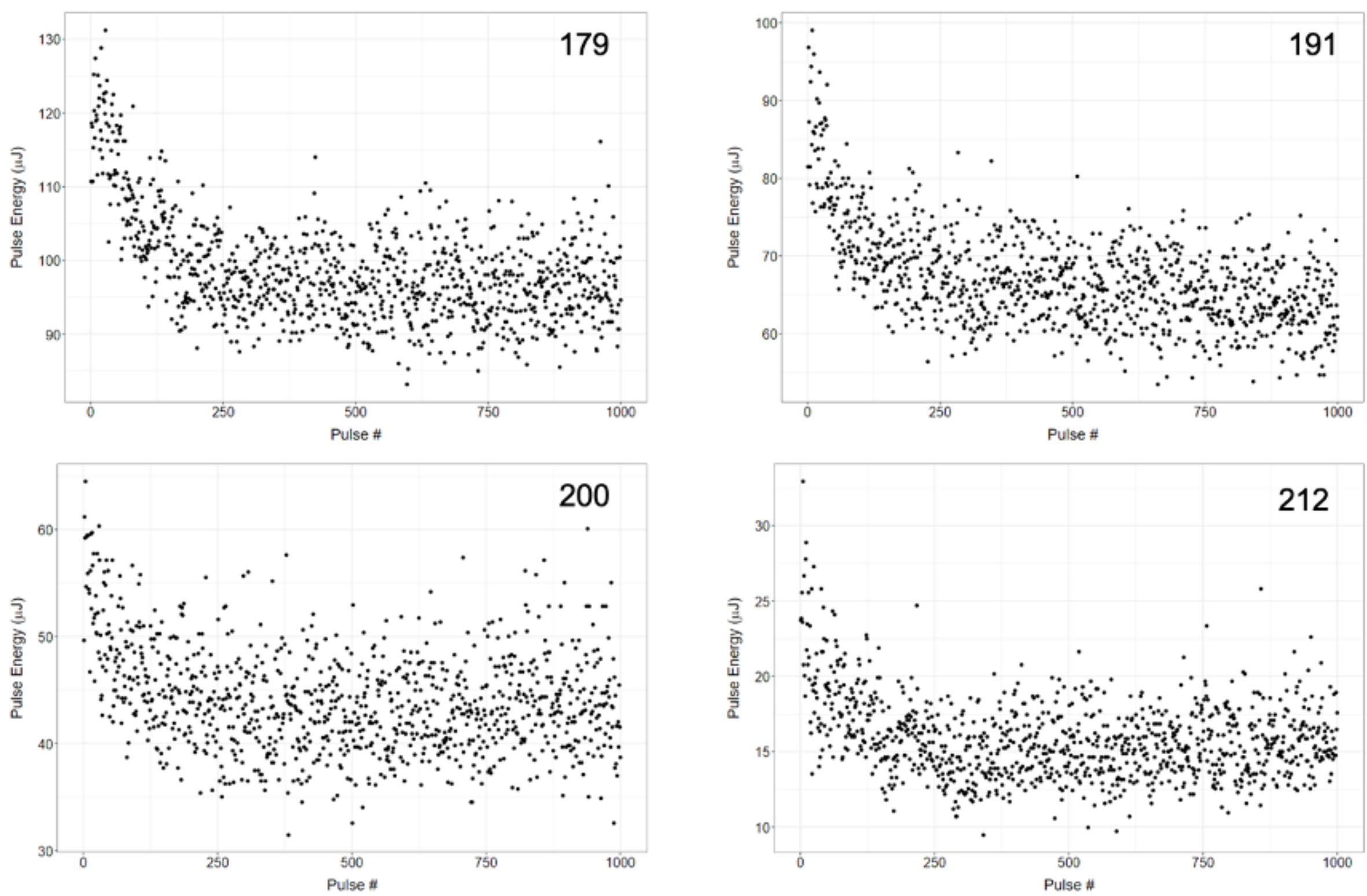

**Supplementary Figure S2.** Laser energy across 1000 pulses during continuous fire at four different Q-switch delay values (179, 191, 200, and 212 µs). Initial laser pulses exhibit higher energy independently from QSD (indicated in the top right corners).

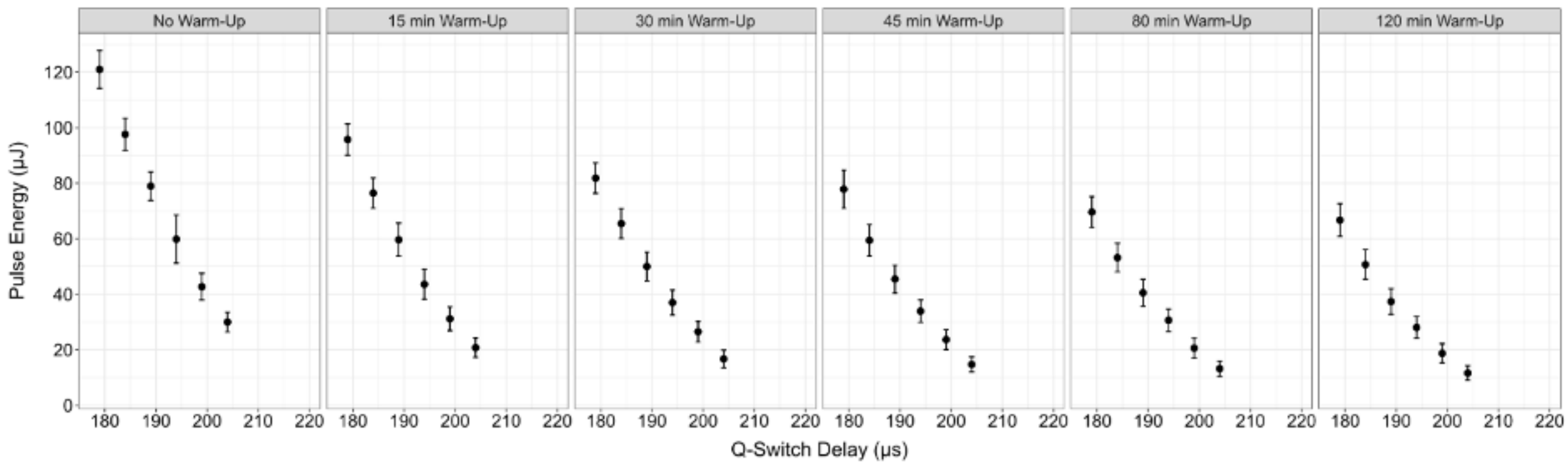

**Supplementary Figure S3.** Optimization of laser warm-up duration. Average laser energy shows a steady decline across QSD values after a warm-up period of continuous fire at QSD = 200 µs, but stabilizes after 30 min particularly at higher delay values (QSD ≥ 200 µs). From left to right: no warm-up, 15 min, 30 min, 45 min, 80 min, and 120 min.

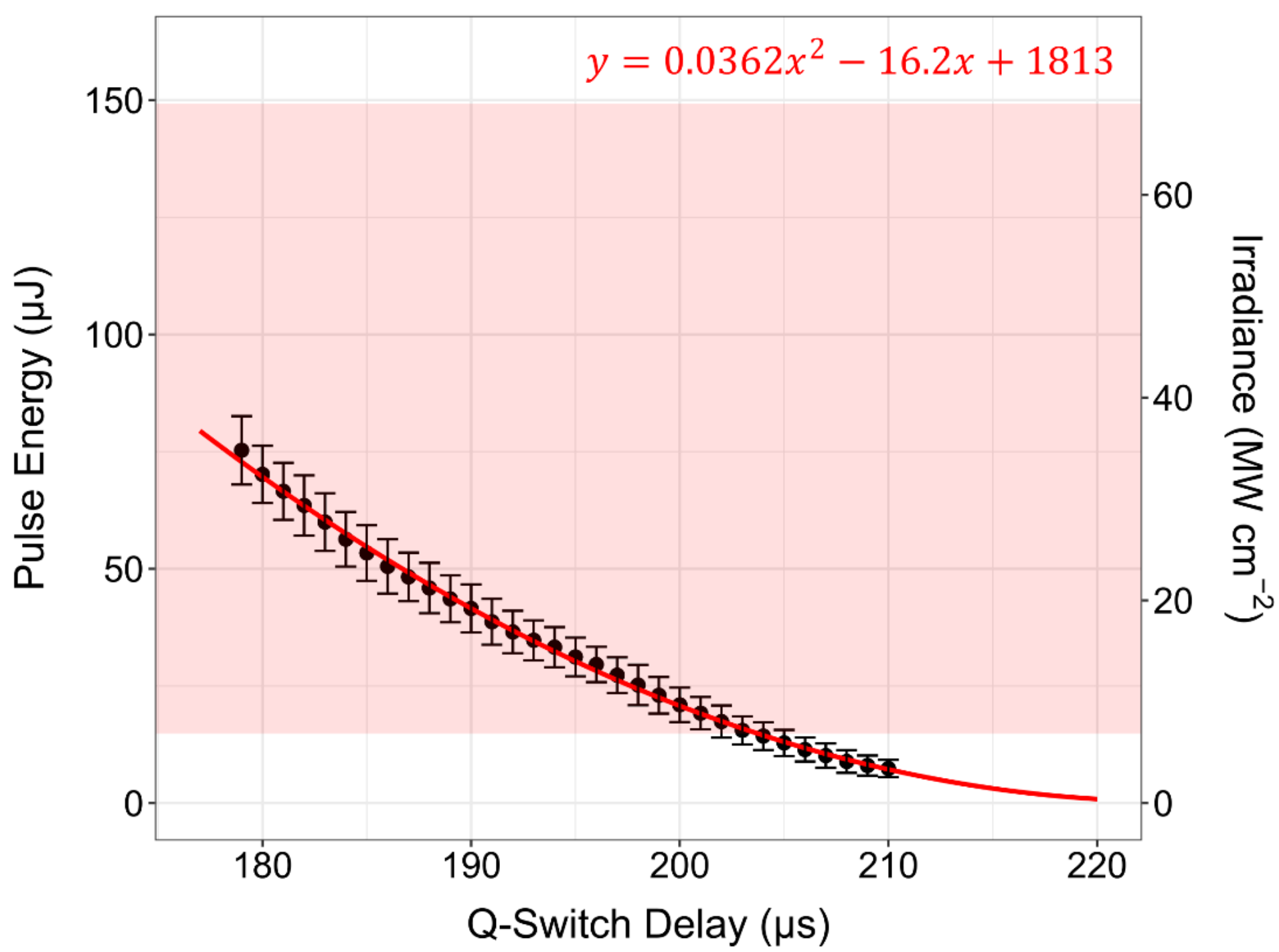

**Supplementary Figure S4:** Calibration of pulse energy (µJ; left y-axis) and irradiance (W cm$^{-2}$; right y-axis) across Q-Switch delay. The curve represents the average of two series of data collected after 30 and 60 minutes of laser warm-up, with removal of the first laser 10 pulses, and fits a second order polynomial regression (adjusted $R^2$ = 0.998). The corresponding MOMA irradiance range is highlighted in red. Data is listed in Supplementary Table S1.

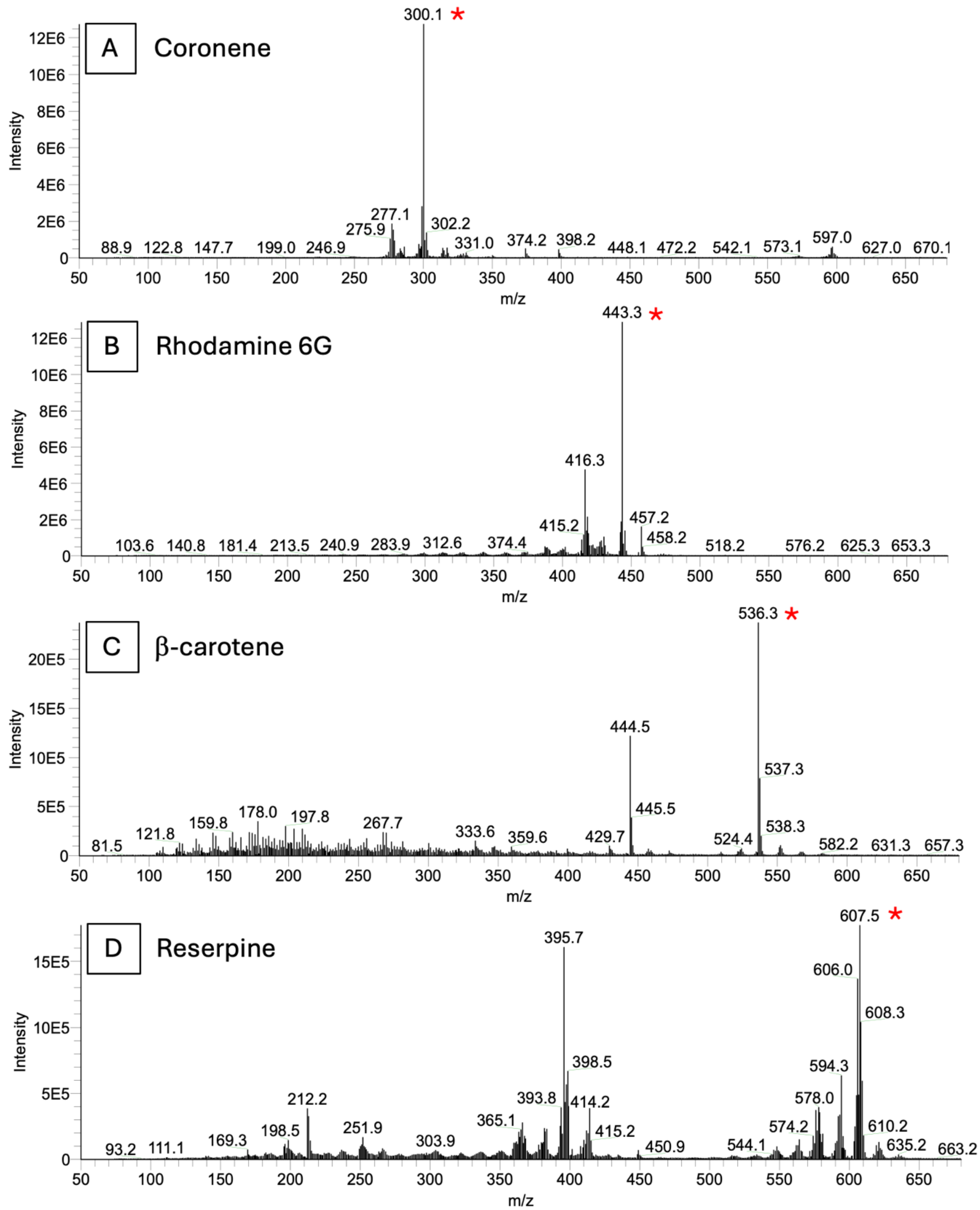

**Supplementary Figure S5.** Mass spectrum of individual standard solutions at 1ug/uL. Despite early fragmentation due to laser energy, parent ions are identified with an asterisk for (A) coronene ($M^+$=300 *m/z*), (B) rhodamine 6G ($M^+$=444 *m/z*), (C) β-carotene ($M^+$=536 *m/z*), and (D) reserpine ($M^+$=608 *m/z*). Laser fluence 0.007 J $cm^{-2}$, comparable to MOMA minimum fluence (See Supplementary Table S1).

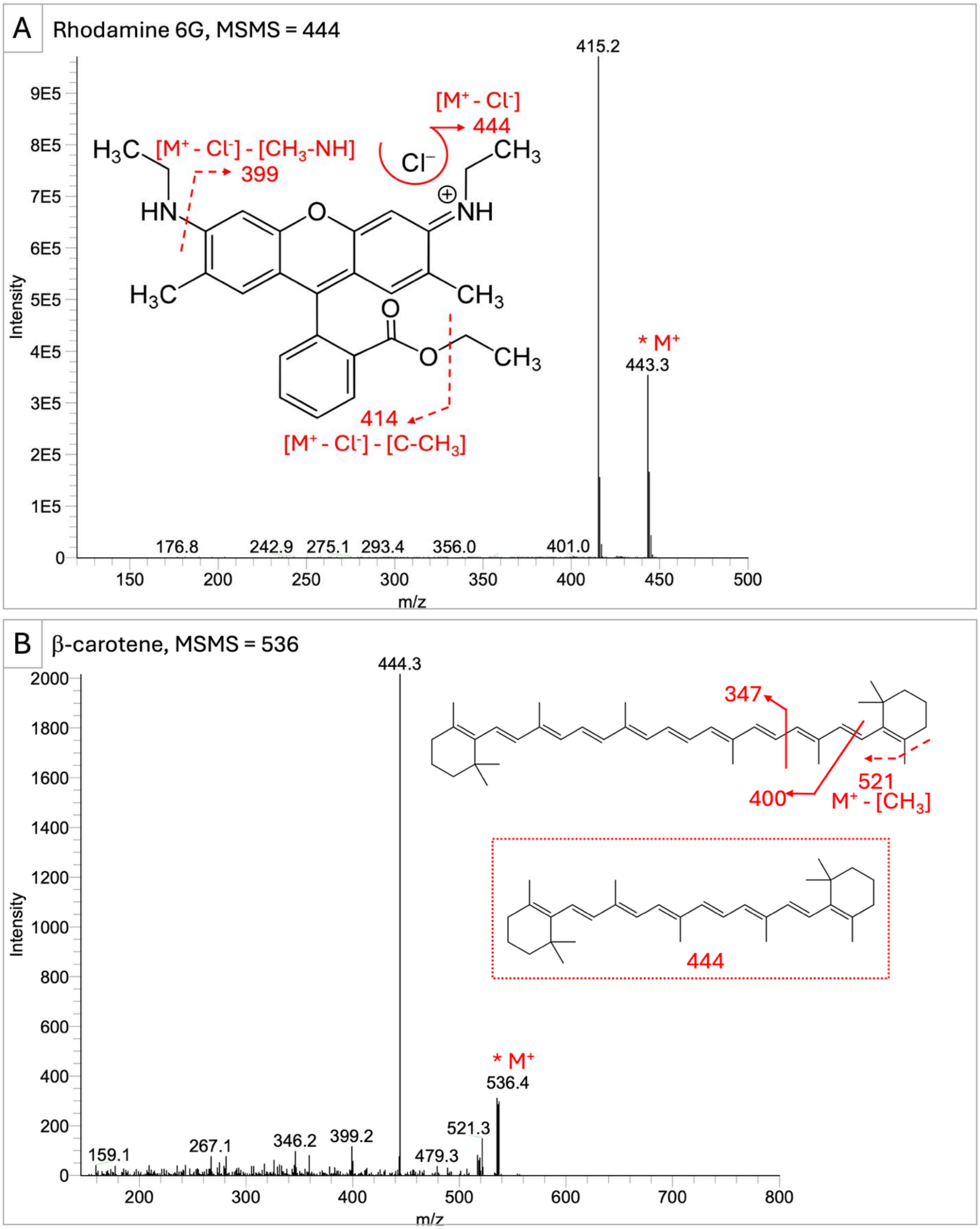

A
Rhodamine 6G, MSMS = 444
415.2
[M+ - Cl-]
444
[M+ - Cl-] - [CH3-NH]
399
Cl−
H3C
CH3
HN
NH
O
H3C
CH3
O
O
CH3
414
[M+ - Cl-] - [C-CH3]
* M+
443.3
176.8
242.9
275.1
293.4
356.0
401.0
9E5
8E5
7E5
6E5
5E5
4E5
3E5
2E5
1E5
0
Intensity
150
200
250
300
350
400
450
500
m/z
B
β-carotene, MSMS = 536
444.3
347
400
521
M+ - [CH3]
444
* M+
536.4
159.1
267.1
346.2
399.2
479.3
521.3
2000
1800
1600
1400
1200
1000
800
600
400
200
0
Intensity
200
300
400
500
600
700
800
m/z

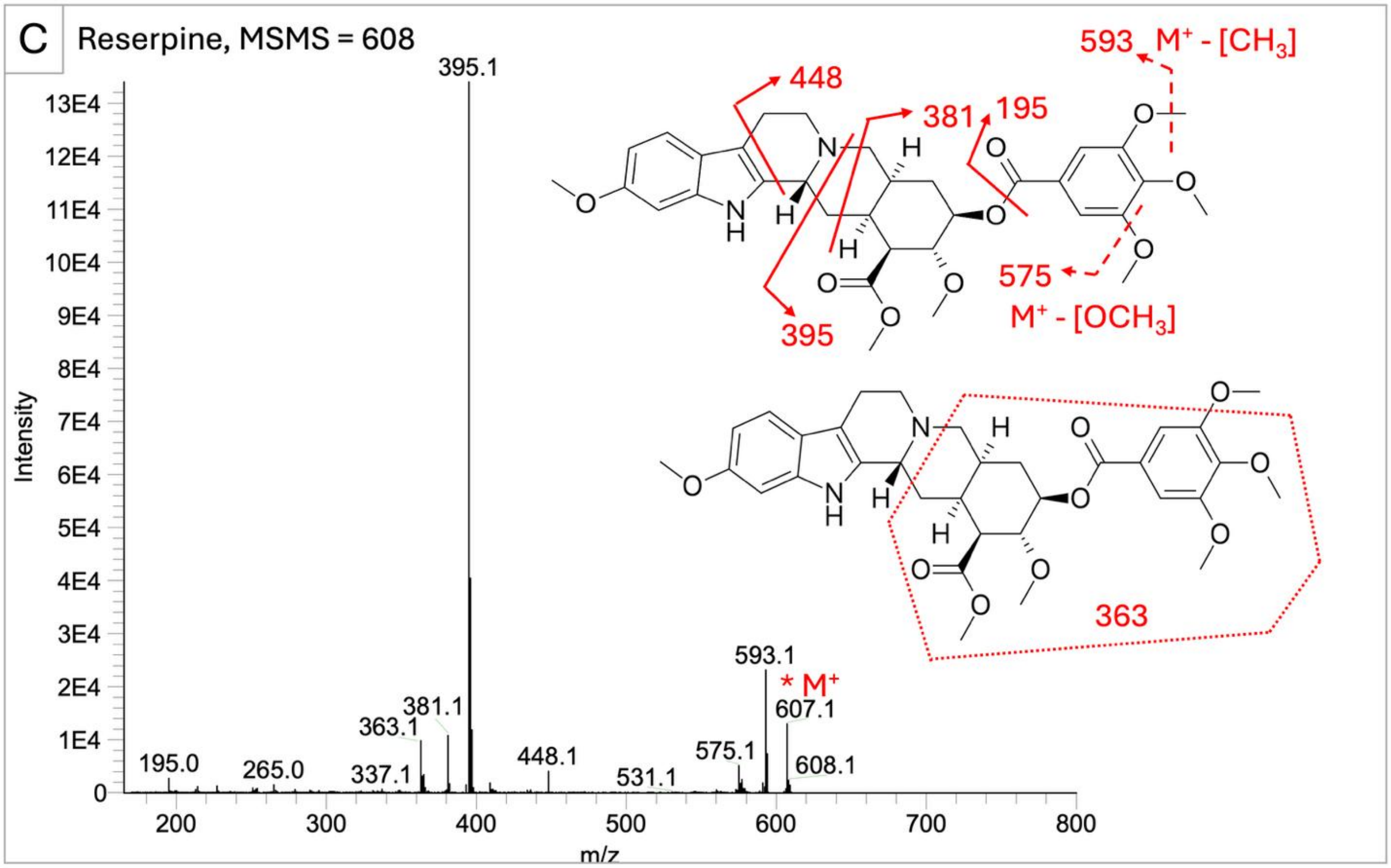

C
Reserpine, MSMS = 608
395.1
593.1
* M+
607.1
608.1
575.1
531.1
448.1
381.1
363.1
337.1
265.0
195.0
Intensity
m/z
448
381
195
593 M+ - [CH3]
575
M+ - [OCH3]
395
363

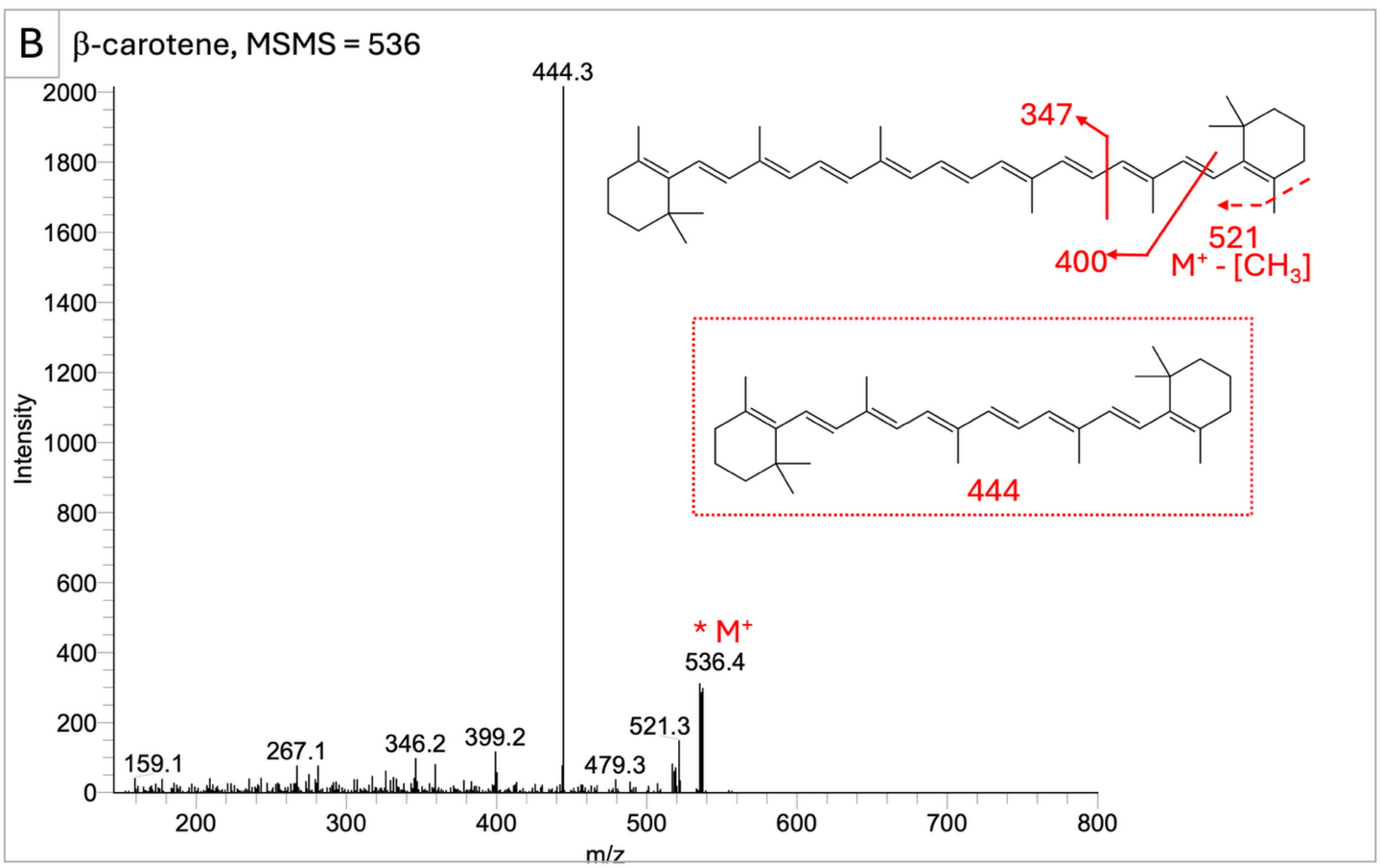

B
β-carotene, MSMS = 536
444.3
* M+
536.4
521.3
479.3
399.2
346.2
267.1
159.1
Intensity
m/z
347
400
521
M+ - [CH3]
444

**Supplementary Figure S6.** Fragmentation patterns of organic standards using MSMS. (A) Rhodamine 6G, CID=45, parent ion mass 444 Da (without the chloride ion), main fragment mass 415 Da, due to the loss of any [C-$CH_3$] group and addition of H. (B) β-carotene, CID=35, parent ion mass 536 Da, main fragment mass 444 Da, hypothetically due to the loss of toluene and recyclization, that can occur while the sample is heated under the laser (Onyewu et al., 1982; Xiao et al., 2015). (C) reserpine, CID=35, parent ion mass 608 Da, main fragment mass 395 Da, due to loss of the three first rings. Fragments in full line indicate exact fragment, and in dotted line, a loss of any group indicated (any [$CH_3$] of the molecule for example). (D) coronene, CID=90, parent ion mass 300 Da, main fragment 273, due to the loss of two external CH groups. For all runs, laser fluence at 0.043 J $cm^{-2}$, comparable to MOMA mid-range fluence (See Supplementary Table S1).

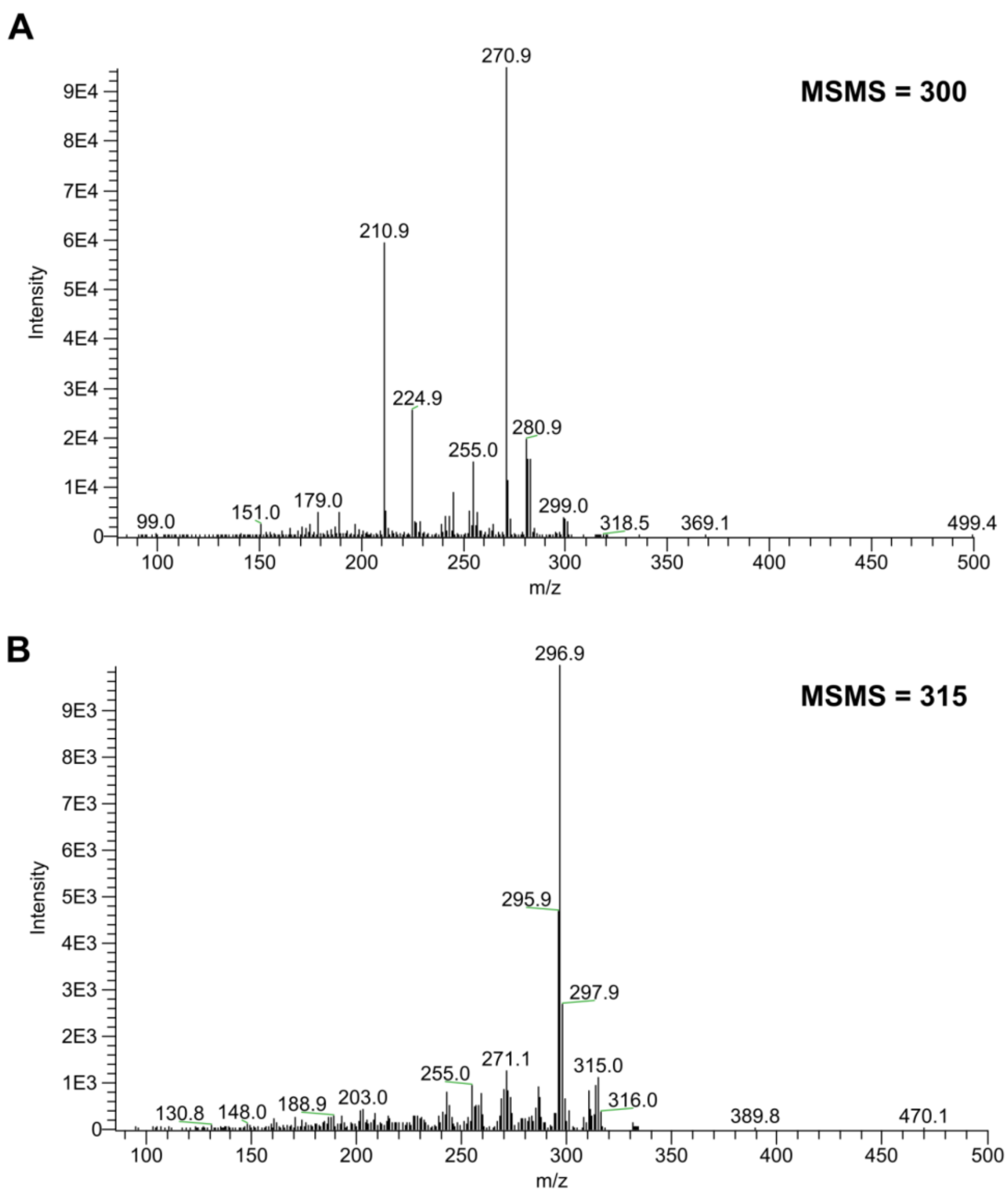

**Supplementary Figure S7.** MSMS spectra of Atacama soil samples. (A) Target parent ion *m/z*=299 from the Surface sample with CID=45. (B) Target parent ion *m/z*=315 from the Surface sample with CID=45. Laser fluence at 0.043 J $cm^{-2}$, comparable to a MOMA mid-range fluence (See Supplementary Table S1).

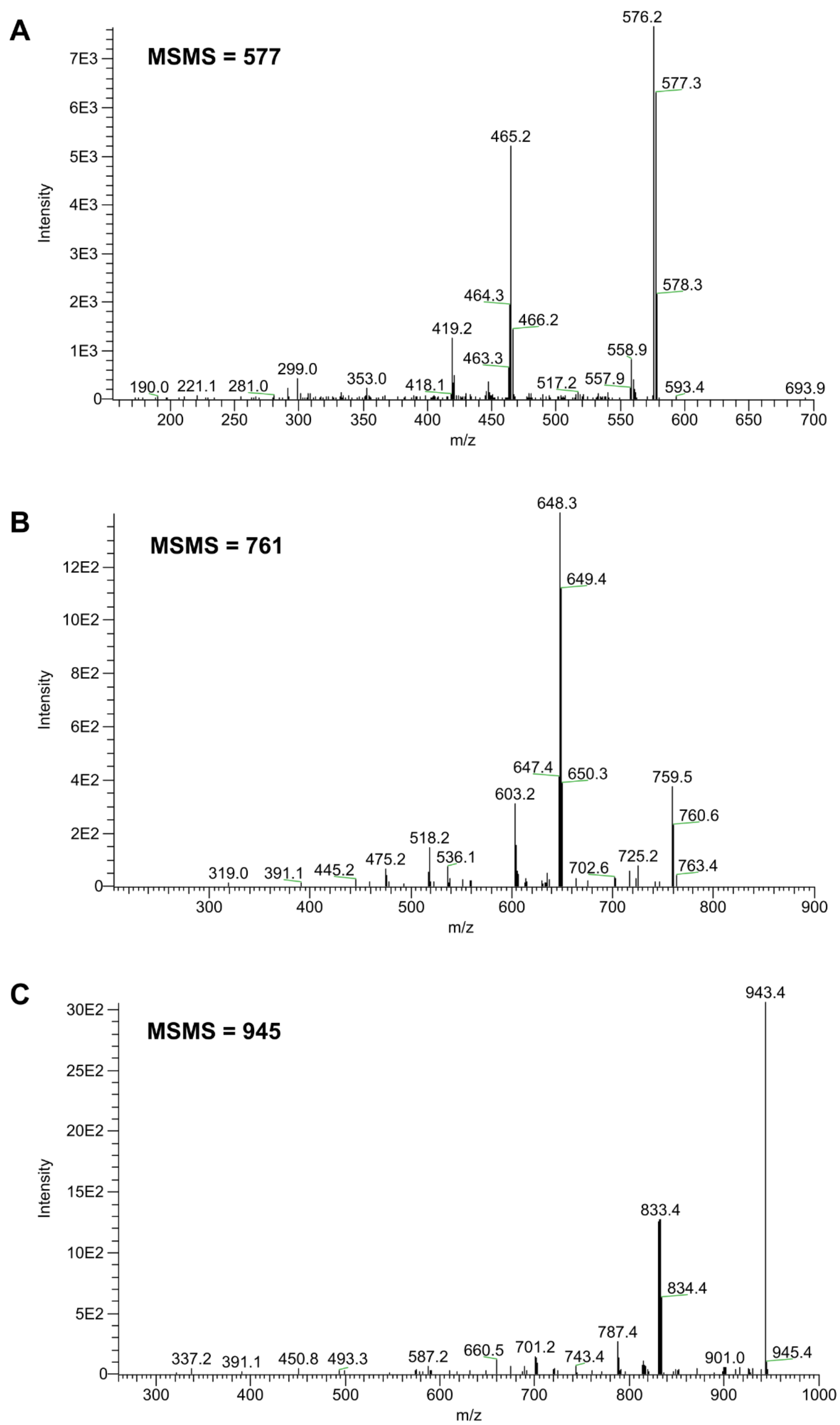

**Supplementary Figure S8.** MSMS spectra of 50 cm depth Atacama soil sample. (A) Target parent ion *m/z*=577 with CID=35. (B) Target parent ion *m/z*=761 with CID=45. (C) Target parent ion *m/z*=945 with CID=45. Laser fluence at 0.043 J $cm^{-2}$, comparable to a MOMA mid-range fluence (See Supplementary Table S1).

***Tables***

**Supplementary Table S1.** Full calibration of laser energy. Two series of data collected after 30 and 60 minutes of laser warm-up are averaged, and the data fits a second order polynomial equation ($y = 0.0362x^2 - 16.2\,x + 1813$), with adjusted $R^2 = 0.998$. The first 10 pulses of each test were removed because of higher laser energy (See Supplementary Figure S2). The MOMA zones indicate the LTQ energies at each QSD that correspond to their equivalent MOMA energy range based on calculated fluence (yellow) and irradiance (red) values. The energies for QSD 211 and above are predicted from the polynomial fit, and the actual MOMA energy range is highlighted in red.

| | LTQ | | | |
|---|---|---|---|---|
| QSD | Energy (μJ) | SD (μJ) | Fluence (J $cm^{-2}$) | Irradiance (MW $cm^{-2}$) |
| 179 | 75.3 | 7.3 | 0.209 | 34.8 |
| 180 | 70.1 | 6.1 | 0.195 | 32.4 |
| 181 | 66.5 | 6.1 | 0.185 | 30.8 |
| 182 | 63.5 | 6.4 | 0.176 | 29.4 |
| 183 | 60.0 | 6.2 | 0.166 | 27.7 |
| 184 | 56.3 | 5.8 | 0.156 | 26.0 |
| 185 | 53.4 | 5.9 | 0.148 | 24.7 |
| 186 | 50.5 | 5.8 | 0.140 | 23.4 |
| 187 | 48.3 | 5.2 | 0.134 | 22.3 |
| 188 | 45.9 | 5.4 | 0.127 | 21.2 |
| 189 | 43.6 | 5.0 | 0.121 | 20.1 |
| 190 | 41.5 | 5.1 | 0.115 | 19.2 |
| 191 | 38.7 | 4.9 | 0.107 | 17.9 |
| 192 | 36.5 | 4.5 | 0.101 | 16.9 |
| 193 | 34.8 | 4.3 | 0.096 | 16.1 |
| 194 | 33.3 | 4.3 | 0.092 | 15.4 |
| 195 | 31.2 | 4.1 | 0.087 | 14.4 |
| 196 | 29.6 | 3.8 | 0.082 | 13.7 |
| 197 | 27.3 | 3.8 | 0.076 | 12.6 |
| 198 | 25.2 | 4.3 | 0.070 | 11.7 |
| 199 | 23.0 | 3.9 | 0.064 | 10.6 |
| 200 | 21.0 | 3.7 | 0.058 | 9.7 |
| 201 | 19.2 | 3.5 | 0.053 | 8.9 |
| 202 | 17.4 | 3.4 | 0.048 | 8.0 |
| 203 | 15.5 | 3.0 | 0.043 | 7.2 |
| 204 | 14.3 | 3.0 | 0.040 | 6.6 |
| 205 | 12.9 | 2.8 | 0.036 | 5.9 |
| 206 | 11.4 | 2.5 | 0.032 | 5.3 |
| 207 | 10.1 | 2.6 | 0.028 | 4.7 |

MOMA zone (irradiance)

MOMA zone (fluence)

| | | | | |
|---|---|---|---|---|
| 208 | 8.9 | 2.4 | 0.025 | 4.1 |
| 209 | 8.0 | 2.1 | 0.022 | 3.7 |
| 210 | 7.4 | 1.8 | 0.020 | 3.4 |
| 211 | 6.2 | | 0.017 | 2.9 |
| 212 | 5.4 | | 0.015 | 2.5 |
| 213 | 4.5 | | 0.013 | 2.1 |
| 214 | 3.8 | | 0.011 | 1.8 |
| 215 | 3.1 | | 0.009 | 1.4 |
| 216 | 2.5 | | 0.007 | 1.2 |
| 217 | 2.0 | | 0.006 | 0.9 |
| 218 | 1.6 | | 0.004 | 0.7 |
| 219 | 1.2 | | 0.003 | 0.5 |
| 220 | 0.9 | | 0.002 | 0.4 |

*cutoff = 10 pulses; n = 580 pulses; model predicted*